\documentstyle[osa,epsfig,manuscript]{revtex}

\newcommand{\la}{\langle}
\newcommand{\ra}{\rangle}

\newcommand{\Ree}{\mbox{$\left< R_e^2 \right>$}}
\newcommand{\Rgg}{\mbox{$\left< R_g^2 \right>$}}

\newcommand{\zee}{\mbox{$\left< z_e^2 \right>$}}

\newcommand{\nuperp}{\mbox{$\nu_\perp$}}

\newcommand{\phitot}{\mbox{$\phi_{t}$}}

\newcommand{\Nav}{\mbox{$\la N \ra $}}
\newcommand{\Ntot}{\mbox{$M_t$}}
\newcommand{\Ng}{\mbox{$\la M_g \ra $}}

\newcommand{\Fav}{\mbox{$\la F \ra $}}
\newcommand{\Hav}{\mbox{$\la H \ra $}}

\newcommand{\Lx}{\mbox{$L_x$}}
\newcommand{\Ly}{\mbox{$L_y$}}
\newcommand{\Lz}{\mbox{$L_z$}}

\begin{document}            
\sloppy        

\title{Formation and Equilibrium Properties of Living Polymer Brushes}
\author{A. Milchev$^1$\thanks{email:milchev@ipchp.ipc.bas.bg}, 
J. P. Wittmer$^2$, and D.P. Landau$^3$}
\address{
$^1$ Institute for Physical Chemistry, Bulgarian Academy of Sciences,
$1113$ Sofia, Bulgaria\\
$^2$ D\'epartment de Physique des Mat\'eriaux, Universit\'e Lyon I \& CNRS,
     69622 Villeurbanne Cedex, France\\
$^3$ Center for Simulational Physics, University of Georgia,
     Athens, Ga. 30602, U.S.A.}
\maketitle
\begin{abstract}   
Polydisperse brushes obtained by {\em reversible} radical chain polymerization
reaction onto a solid substrate with surface-attached initiators, are studied 
by means of an off-lattice Monte Carlo algorithm of living polymers (LP). 
Various properties of such brushes, like the average chain length and the 
conformational orientation of the polymers, or the force exerted by the 
brush on the opposite container wall, reveal power-law dependence on the
relevant parameters.
The observed molecular weight distribution (MWD) of the grafted LP decays
much more slowly than the corresponding LP bulk system due to the gradient 
of the monomer density within the dense pseudo-brush 
which favors longer chains. 
Both MWD and the density profiles of grafted polymers and chain ends
are well fitted by {\em effective} power laws whereby
the different exponents turn out to be mutually self-consistent
for a pseudo-brush in the strong-stretching regime. 
The specific values are, 
however, inconsistent with a standard self-consistent field theory of 
pseudo-brushes which predicts a much softer 
mushroom-like layer. 
\end{abstract} 

\newpage

\section{Introduction}
\label{sec:Intro}

Due to their potential for practical applications\cite{HTL92}, polymer brushes 
(layers of linear polymer chains end-grafted to a surface) have been the topics 
of vivid scientific interest. Most of the experimental
and theoretical studies so far, have been devoted to
monodisperse brushes of polymer chains in which the reaction of
polymerization has been terminated \cite{BLOBBRUSH,SCFBRUSH,NS98}. 
The conventional way of forming such brushes with sufficiently high density, 
whereby pre-existing polymer chains with functionalized groups on the one end 
are adsorbed on a surface, turns out to be extremely slow and 
inefficient\cite{Slowprocess}. 
Thus, the attachment of polymer molecules, once the surface is significantly 
covered, is heavily suppressed by kinetic hindrance.

An effective way to achieve sufficiently high grafting densities is to grow
the layer {\em in situ}, monomer by monomer, from a functionalized seed
carrying polymerization initiators\cite{Ruehe}. 
Theoretically, the growth of polymer chains from a surface was studied
recently by Wittmer et al.\cite{WCJT96}, combining elements of
Diffusion-Limited Aggregation (DLA) \cite{DLA} with the theory of 
polydisperse strongly stretched polymer brushes \cite{BLOBBRUSH,SCFBRUSH,Guis}.
Generalizing the "needle growth" problem\cite{Needles}, they view
the formation of the brush as a particular case of 
Diffusion-Limited Aggregation Without Branching (DLAWB):
An infinitesimally small incident flux of matter feeds the growth 
so that the polymer coils (in the good solvent regime 
\cite{Degennesbook,DJbook}) 
can relax their structure in response to excluded-volume interactions with 
their neighbors while new free monomers are {\em irreversibly} 
attached at the active ends of the chains. 
This approach predicts strong differences between polymer molecules grown on
surfaces and polymers formed in the bulk. The polydispersity
of the former is much higher compared to that for the same reaction in 
the solution because in a brush the longer chains are more efficient in 
adding new monomers than the denser short chains which have to compete
for fewer "random walkers". 
Below we briefly sketch some of the main predictions of this approach
(see Sec.\ref{subsec:DLAWB}).

\begin{figure}
\centerline{\epsfig{file=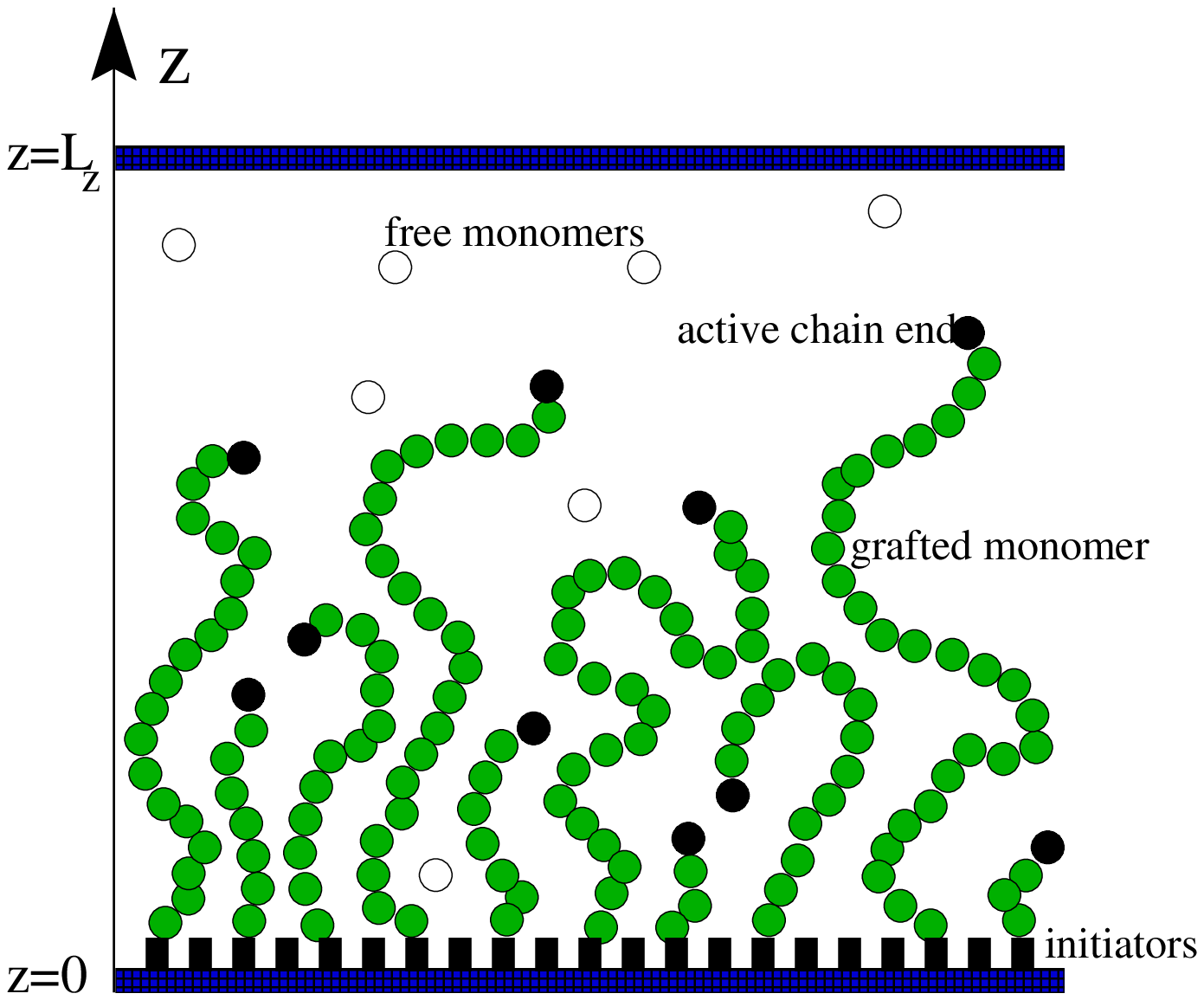,width=120mm,height=100mm}}
\vspace*{1cm}
\caption{Sketch of our model of a living polymer (LP) brush. 
Polymer chains grow {\em reversibly} from the (fully activated) starters 
(dark squares) fixed on the (impenetrable) bottom wall. 
They are in thermal equilibrium with a reservoir of free monomers
(open spheres). We suppose that both the scission energy $J$ and the 
activation barrier are independent of the monomer position
(along the chain contour as well as in space) and density. 
Desorption events occur only at the active chain ends (dark spheres)
and chains are not allowed break. Branching of chains is forbidden as well. 
The brush is put into a container with an opposite wall at $z=\Lz$.
The total number of particles \Ntot\ is conserved.
\label{fig:sketchPB}}\
\end{figure}

In the present work we try to extend the investigation of {\em in situ} 
grown polymer brushes, focusing on their properties under conditions of 
chemical equilibrium between the polymers and their respective monomers
as sketched in Fig.\ref{fig:sketchPB}.
To this end a flat impenetrable surface is densely covered with 
(fully activated) initiators. These sites serve as anchoring points for 
linear unbranched chains, so-called "Living Polymers" (LP) \cite{Greer}, 
which can grow by {\em reversible} polymerization until an equilibrium 
with the ambient phase of free monomers is established. 
We suppose that both the scission energy $J$ and the
activation barrier $B$ are independent of the monomer position
(along the chain contour as well as in space) and density.
Desorption events occur only at the active chain ends.
In contrast to the systems of "Giant Micelles" (GM) considered elsewhere
\cite{CC90,WMC98}, the grafted LP chains are not allowed to break
and there are no freely floating LP in the bulk. 
The LP brush is grafted at the bottom (at $z=0$) of a long
container with an additional inert wall at the top ($z=L_z$),
i.e. we consider a canonical ensemble and the total number of grafted
and free monomers $\Ntot=N_g+N_f $ is conserved.

Although this system is qualitatively different from the case of 
irreversible DLAWB growth\cite{WCJT96}, some properties that we have 
investigated by means of an off-lattice Monte Carlo algorithm
are found to match surprisingly well. Thus, the
longer chains are again favored (compared to corresponding bulk systems)
due to the gradient of the density profile of grafted monomers. Hence, 
the polydispersity again becomes very large and we observe (surprisingly) 
virtually the {\em same} power law behavior for the MWD $c(N)$ 
as reported by Wittmer et al. \cite{WCJT96}. 
This finding pertains to the established density profiles of 
aggregated monomers $\phi(z)$ and active chain ends $\rho(z)$ 
as function of their distance $z$ from the grafting surface as well.

We will first sketch in Section \ref{sec:Theory} the elements of a 
theoretical description of a dense LP brush. 
The computational algorithm is then outlined in Section \ref{sec:Algo}.
The preparation of equilibrium configurations and the kinetics of layer growth
until the onset of equilibrium is briefly examined in Section 
\ref{sec:Kinetics}, and the main results of this paper on  the static 
properties of LP brushes are presented in the subsequent Section 
\ref{sec:Static}.  We examine the influence of total monomer concentration 
\phitot\ and grafting density $1/d^2$ of the initiators on the density 
profiles, the average degree of polymerization \Nav, 
and on the conformational properties of the polymer chains in the brush.
In order to make connection with possible experiments using
surface force machines, the force \Fav, exerted by the brush on the 
opposing wall of the container, is investigated.
In Section \ref{sec:Discussion} we briefly discuss the computational results 
in view of the self-consistent field theory presented in Section \ref{sec:Theory},
and we conclude with a short Summary of the main results of this work.

\section{Analytical Considerations}
\label{sec:Theory}

\subsection{Mean-Field considerations}
\label{subsec:MFassumption}

In contrast to the DLAWB brush discussed by Wittmer at al. \cite{WCJT96},
where the polymerization was assumed to be {\em irreversible},
we consider here a grafted polymer layer with annealed mass distribution
in thermal equilibrium with ambient free monomers.
Obviously, the task is to minimize the total free energy of the system 
\cite{Godefroid}
which is in general a functional of the free monomer distribution $u(z)$,
of the density profile of the grafted monomers $\phi(z)$
and of the MWD $c(N)$.
At low enough free monomer concentration we may write this
total free energy as the sum of the free energy of the brush
$F_{brush}[\phi(z),c(N)]$ and the free energy of the free monomers
$F_{free}[u(z),\phi(z)]$.
The latter contains the usual entropy of mixing term as well
as an excess chemical potential due to the interaction with the
dense LP layer formed around $z=0$. 
From minimization of both terms, subject to the constraint of
fixed total particle number \Ntot, 
the mean number of monomers aggregated in the layer \Ng\ may in principle 
be calculated as a function of \Ntot\ and the interaction parameters. 
Being only interested in the functional form of the MWD and the
layer profiles, we merely state that such an 
"equation of state" can be found and an arbitrary number of monomers $M_g$
can be bound per unit surface (within some obvious limits).
Hence, (ignoring fluctuations of $M_g$) we decouple the problem in two parts, 
one for the free monomers with constant chemical potential 
$\mu(\Ntot)=\log(u(z))+\mu_{ex}(\phi(z))$
and one for the $M_g$ monomers in the layer
at same chemical potential.

The simplified task is now to minimize the free energy of the brush
subject to the constraint of $M_g$ monomers per layer. It is natural to
write this free energy in a Flory-Huggins like manner as 
\begin{equation}
F_{brush}[\phi(z),c(N)] =  \sum_{N=1}^{\infty}
c(N) \left( \log(c(N)) + \mu_1 N + F_{chain}[\phi(z),c(N),N] \right).
\label{eq:Fbrush}
\label{eq:MFhyp}
\end{equation}
Note that all densities are taken per unit surface
and that the Boltzmann factor is set equal to one.
The first term is the entropy of mixing for the grafted chains,
the second entails the usual Lagrange multiplier\cite{foot1} for the
conserved number of aggregated monomers $M_g$ (see above). 
The last term is the free energy of a reference chain in
the self-consistent density profile created by neighboring chains,
which in general is a function of chain length $N$ and
a functional of both $\phi(z)$ and $c(N)$.
The above mean-field equation assumes in particular
that the length of a given reference chain is not
correlated with the chain lengths of its neighbors.
We note that all contributions to $F_{chain}[\phi(z),c(N),N]$ 
which are linear in chain length (even if dependent on $\phi(z)$ and $c(N)$)
can be incorporated in the Lagrange multiplier term $\mu_1 N$ and 
are hence irrelevant. 

Before we continue to tackle the full self-consistent field problem
in Sec.\ref{subsec:SCF_LPBRUSH} within the strong-stretching assumption of 
polymer brush theory (Sec.\ref{subsec:SSH}) we need to recap
some elements of the theory of polymers and polymer layers.
In the reminder of this subsection we will consider some simple
cases of the general problem formulated above 
where the free energy of the reference chain is a function of $N$ only. 
Hence, one readily obtains the equilibrium MWD from the functional derivation 
of $F_{brush}[c(N)]$ with regard to $c(N)$ the equilibrium MWD
\begin{eqnarray}
c(N) & \propto & \exp(-\mu_1 N - F_{chain}(N)) 
\label{eq:cN}.
\end{eqnarray}

Two particular simple cases where this applies 
are a forests of needles growing vertically 
on a flat substrate and Gaussian polymer chains 
(without excluded volume interaction) fixed at a surface.
In both cases the free energy per chain is linear in $N$ and the
MWD is simply $c(N) \propto \exp(-N/\Nav)$.

We can now also address the problem of dilute non-overlapping LP
(so-called {\em mushrooms}) in good solvent fixed on the surface.
If the surface is {\em penetrable} the free energy of the reference
chain is the same as for dilute LP in the bulk, i.e.
$F_{chain}(N)=(\gamma-1)\log(N)$ where $\gamma=1.165 > 1$ 
due to long-range excluded volume correlations along the chain
\cite{Degennesbook,DJbook}.
This gives rise to a Schultz-Flory distribution and was recently
confirmed computationally\cite{WMC98,MWL99b}.
The situation is slightly different for LP on an {\em impenetrable} surface
which reduces the partition function and effectively repels the chains
\cite{DJbook}.
In this case one expects $F_{chain} = -\tau \log(N)$ with the
exponent $\tau=1-\gamma_s$ where $\gamma_s \approx 0.65 < 1$. 
Hence, one expects to find the weakly singular MWD
\begin{equation}
c(N) \propto N^{-\tau} \exp(-\mu N).
\label{eq:cNmush}
\end{equation}
We will re-address this issue\cite{foot2} in Sec.\ref{subsec:MWD}.

\subsection{Weakly-stretched mushroom-like layer}
\label{subsec:WSH}

If the total amount of monomers $M_g$ is now further increased
and the mushrooms start to overlap, the description of the
equilibrium LP layer becomes more complex.
One expects the excluded volume interactions between the monomers
to favor longer chains which can explore more dilute regions of the layer.
It is hence conceivable that the MWD is not a pure exponential,
but a power law (plus cut-off). 

There are two possible extreme scenarios or architectures which are 
relatively easy to compute. One is to suppose that the LP chains
strongly overlap and form a strongly stretched LP brush.
We will consider this in the next subsection.
A second natural guess is to speculate
that the excluded volume interaction is extremely strong (compared to the
entropic terms in $F_{brush}$) and that, therefore, the layer attempts 
to reduce chain overlap as far as possible \cite{Godefroid}.

A geometrical construction for this limiting case is
a self-similar mushroom-like structure with $\xi(z) \propto z$.
Hence, every LP chain is essentially contained within its own
excluded volume blob of size $\xi$ 
containing $g \propto \xi^{1/\nu}$ monomers.
Here, $\nu \approx 3/5$ denotes the Flory exponent\cite{Degennesbook} 
in dimension $D=3$. This forces the density of monomers to scale like 
$\phi(z) \propto g(z)/\xi(z)^3 \propto z^{-\alpha}$
with $\alpha=3-1/\nu \approx 4/3$ and 
the density of ends like the blob density 
$\rho(z) \propto 1/\xi(z)^3 \propto z^{-\beta}$ with $\beta=3$.
From this the MWD is easily found using $c(N) dN = \rho(z) dz$.
This yields $c(N)\propto N^{-\tau}$ with $\tau=1+2\nu \approx 11/5$.
Note that because $\alpha > 1$ most of the layer mass is to be found
at the lower cut-off should this scenario apply, 
i.e. if it indeed minimizes the total brush free energy.
 
\subsection{Strongly Stretched Alexander-De Gennes brush}
\label{subsec:SSH}

In this subsection we assume
that the equilibrium LP layer is described by a strongly-stretched
pseudo-brush of Alexander-De Gennes type\cite{BLOBBRUSH},
i.e. a compact layer of blobs of size $\xi(z)\propto \phi(z)^{-1/(3\nu-1)}$.
Coarse-grained on distances larger than the blob size $\xi$,
the chain are described by "classical trajectories" 
\begin{equation}
n(z) = \int_{0}^{z} g(z') dz'/\xi(z')
\label{eq:trajectory}
\end{equation} 
where the monomer index $n$ is counted from the grafted initiator ($n=1$).
Within this geometrical construction the chains are strongly stretched
on distances larger than $\xi$.
Hence, the number of chains per unit surface at $z$ is given by
\begin{equation}
\int_z^H \rho(z') dz' = \int_{N}^{N_H} c(N') dN' \propto \phi^{1/\epsilon}.
\label{eq:countends}
\end{equation}
with $\epsilon=2\nu/(3\nu-1)\approx 2/3$, $H$ denoting the upper
edge of the pile of blobs while $N_H = n(z=H)$ stands for the upper cut-off
of the layer so that the ends of all chains passing
through an arbitrary plane at $z$ stay beyond that plane \cite{Godefroid}.
For the last term in eq.(\ref{eq:countends}) we have used the fact that 
the number of chains equals the number of blobs at $z$ 
(which itself scales like $1/\xi^2$)
and the connection between $\phi$ and $\xi$ imposed by the
fractal dilute structure within the blob.

Anticipating our computational results on LP brushes (Sec.\ref{sec:Static}),
we use power-law functions to express 
$\phi(z) \propto z^{-\alpha}$,
$\rho(z) \propto z^{-\beta}$,
$c(N) \propto N^{-\tau}$ and
$z(n) \propto n^{-\nuperp}$,
as in the case of the weak-stretching limit (Sec.\ref{subsec:WSH}).
Plugging this into eqs. (\ref{eq:trajectory})-(\ref{eq:countends}), we obtain
three independent equations for our four exponents \cite{Godefroid}. We express 
them in terms of $\tau$, the exponent of the MWD: 
\begin{eqnarray}
\alpha & = & \frac{(3\nu-1)(\tau-1)}{1+\nu-(1-\nu)\tau} 
\label{eq:alpha_tau} \\
\nuperp & = & 1/(1+\alpha (1-\nu)/(3\nu-1)) =
(1+\nu-(1-\nu) \tau)/2\nu \label{eq:nuperp_tau} \\ 
\beta & = & \alpha/\epsilon+ 1 = 
\frac{(3\nu-1)\tau - \nu + 1}{1+\nu-(1-\nu)\tau} 
\label{eq:beta_tau} 
\end{eqnarray}
The second of these equations results from the "trajectory" 
eq.(\ref{eq:trajectory}),
the third from equating the first and last term in eq.(\ref{eq:countends}). 

One possible solution of the above three relations is --- and this
might at first sight appear surprising --- the set of exponents derived
above for the weakly-stretched mushroom-like layer (with $\nuperp=\nu$).
This just means that there is a smooth matching between the regimes
of strong and weak stretching --- at least for the exponents \cite{Godefroid}.

Another possible set of exponents which is consistent with the 
three strong-stretching relations arises for a brush formed by
irreversible diffusion-limited aggregation.
 
\subsection{Diffusion Limited Aggregation without Branching}
\label{subsec:DLAWB}

Here we briefly recall the results \cite{WCJT96} derived
by combining a mean-field treatment of the diffusive growth 
with a scaling theory of the growing polymers.
The theory considers the case of slow growth due to irreversible 
polymerization which is supplied by a constant infinitesimal flux of incoming 
monomers. 
The continuum field description of the growth is based on the above
strong-stretching equations and, in addition to this, to two
standard equations for aggregation processes assuming 
a second-order process for the reaction and
mass conservation in the "adiabatic approximation"\cite{DLA}.
It is possible to solve these non-linear partial differential
equations by means of a power law scaling ansatz \cite{WCJT96}.
To be consistent, the Laplacian field of incoming random walkers, 
requires one additional equation between the exponents, $\beta=2$,
which fixes the above set of exponents in a unique way:
$\tau=7/4$, $\nuperp=3/4$, $\beta=2$ and $\alpha=2/3$.
Hence, the layer is much more brush-like with $\xi(z)\propto \sqrt{z}$
than the fluffy mushroom-like layer considered above.
These exponents, as well as the related scaling relations, have
been reproduced by means of MC simulations showing the
DLAWB to be marginally mean-field like in 3D
in contrast to the standard DLA problem \cite{DLA}.

\subsection{Dense LP brush}
\label{subsec:SCF_LPBRUSH}

Finally we consider the original problem of minimizing the 
free energy of a dense brush $F_{brush}[\phi(z),c(N)]$
in the limit of high coverage $M_g$. 
Following a standard SCF procedure\cite{SCFBRUSH}, we integrate out
the $\phi(z)$ dependence of the free energy \cite{Godefroid} and assume 
the free energy for a given MWD $c(N)$ to be well described by the 
Alexander-De Gennes blob construction.
Hence, we neglect the logarithmic terms not taken into
account in this steepest-decent approximation.

The free energy of a reference chain still depends
on the MWD and the mass distribution, obtained
by functional derivation with respect to $c(N)$, 
will be in general different from eq.(\ref{eq:cN}).
For the minimization it is more convenient to use 
the total excess energy in the layer due to the chain
interaction $F_{ev}[c(N)] = \sum_N c(N) F_{chain}[c(N),N]$ 
rather than $F_{chain}[c(N),N]$.
We assume that $F_{ev}[c(N)]$ is well approximated by the
total number of blobs within the layer
\begin{equation}
F_{ev}[c(N)] = \int_z^H dz/\xi(z)^3 \propto 
\int_{n=0}^{N_H} dn \phi(n)^{(2\nu+1)/(3\nu-1)}
\propto \int_{n=0}^{N_H} dn \left( \int_{n'>n}^{N_H} dn' c(n') \right)^\kappa 
\label{eq:Fev}
\end{equation}
where $\kappa=1+1/2\nu\approx 11/6$.
For the last equation above we have used
$\phi(n)=\left( \int_{n'>n}^{N_H} dn' c(n') \right)^{(3\nu-1)/2\nu}$
from the strong-stretching relation eq.(\ref{eq:countends}).
A careful computation of the functional derivation of the total free energy of
the layer, eq.(\ref{eq:Fbrush}), generalizing eq.(\ref{eq:cN}), yields 
\begin{equation}
0 = \frac{\delta F_{brush}}{\delta c(N)} =
\log(c(N)) + \mu_1 N + 
\int_0^N dn \left(  \int_n^{N_H} dn' c(n') \right)^{\kappa-1}
\label{eq:cNbrush}
\end{equation}
where we have left out irrelevant constants and prefactors.
The second term on the right side describes the (trivial) exponential cut-off. 
For thick layers (large $M_g$) we examine again
the asymptotic power law behavior at $z \ll H$. 

Anticipating the result of Fig.\ref{fig:mwd} we may look
for a solution of eq.(\ref{eq:cNbrush}) with $c(N)\propto N^{-\tau}$,
that is, we require the integrand in eq.(\ref{eq:cNbrush}) to be $\propto 1/n$.
This argument would not be changed if additional logarithmic terms
$\propto \log(N)$ are added by hand in eq.(\ref{eq:cNbrush}) \cite{Godefroid}.
It is simple to check that this yields $\tau=1/(\kappa-1)+1=1+2\nu=11/5$
which is the same exponent found previously in Sec.\ref{subsec:WSH} for
the  mushroom-like layer with $\xi(z)\propto z$.  
Within the strong-stretching assumption eqs.(\ref{eq:trajectory}) to (\ref{eq:countends})
this yields the exponents $\nuperp=\nu=3/5$, $\beta=3$ and $\alpha=2/3$.

This is obviously a peculiar result in that by using at various
points the strong-stretching assumption we eventually obtain a fluffy layer. 
However, as was stressed at the end of Sec.\ref{subsec:SSH},
both physical pictures match smoothly and one might still regard 
this results as marginally self-consistent.

\section{MODEL AND METHOD}
\label{sec:Algo}

Since we are not attempting to describe the properties of particular 
system (such as polystyrene, etc.) but rather wish to contribute to the 
general understanding of universal properties of {\em in situ} grown 
LP brushes, we may use an algorithm in which each chain consists of 
coarse-grained monomers connected bonds representing at least, say, 
$3 - 6$ chemical bonds along the backbone of a polymer chain. 

In the present investigation we have harnessed the coarse-grained 
bead-spring algorithm for polymer chains\cite{AMKB} already applied 
successfully for systems of GM in ref.\cite{MWL99b}.
Our description can therefore be brief. 
The main difference with regard to the previous study is that 
--- following the model definition sketched in Fig.\ref{fig:sketchPB} ---
the monomers may attach or dissociate reversibly only from end monomers of 
grafted LP chains.

Each bond is described by a shifted {\sc FENE} potential where a bond of length $l$
has a maximum at $l_{max}=1$
\begin{eqnarray}
U_{FENE}(l)=-K(l_{max}-l_0)^2\ln \left [ 1-
\left (\frac{l-l_0}{l_{max}-l_0}\right )^2 \right ] - J
\label{eq:FENE}
\end{eqnarray}
where $J$ corresponds to the constant scission energy introduced 
in the Introduction.
Note that \mbox{$U_{FENE}(l=l_0)=-J$} and that
$U_{FENE}$ near its minimum at $l_0$ is harmonic, with $K$ being the spring
constant, and the potential diverges to infinity both when $l\rightarrow 
l_{max}$ and $l\rightarrow l_{min}=2l_0-l_{max}$. 
Following ref.\cite{AMKB} we choose the parameters $l_{max}-l_0=l_0-l_{min}=0.3$ 
and $K/T=20$, $T$ being the absolute temperature.  
The units are such that the Boltzmann's constant $k_B = 1$.

The non-bonded interaction between effective monomers is described by 
a Morse-type potential, $r$ being the distance between the beads 
\begin{eqnarray}
U_M(r)= \exp[-2a(r-r_{min})]-2\exp[-a(r-r_{min})]
\label{eq:Morse}
\end{eqnarray}
with parameters $a=24$ and $r_{min}=0.8$. 
Then the $\Theta$-temperature of the coil-globule transition 
for our model is $\Theta \approx 0.62$ so that at $T=1$ we work under good
solvent conditions\cite{MPB}.

The model can be simulated fairly efficiently with a dynamic MC algorithm,
as described previously\cite{MPB,AMKB}. The trial update involves
choosing a monomeric unit at random and attempting to displace it randomly
by displacements $\Delta x, \Delta y, \Delta z$ chosen uniformly from the
interval $-0.5\leq \Delta x, \Delta y, \Delta z\leq 0.5$. Moves are
then accepted according to the Metropolis criterion and one Monte Carlo 
step (MCS) 
involves as many attempted moves as there are monomers in the system. 
In addition, during each MCS as many bonds as there are initiators in the
system are chosen at random at the active ends of the grafted chains, 
and an attempt is made to break them according to the Metropolis algorithm.  
Attempts are also made to create new bonds between the end monomers and  
free monomers within the potential range of $U_M$  (i.e. a new bond with 
energy $U_{FENE}$)\cite{MWL99b}. 

In order to keep the system in equilibrium with the ambient phase of single 
free monomers and prevent the longer polymer chains from touching the top of 
the container we have used a rather low value of the bond energy 
$J = 2$.
The lattice constant $d$ of the square grid of activated initiators is taken 
as a rule as $d=1$ for the case of a dense brush and, as a special case of a 
loose mushroom-like layer, $d=4$.
Typically, boxes of size $L_x\times L_y\times L_z$ and periodic boundary 
conditions in $x-$ and $y-$directions have been used in the simulations
with $L_x=L_y=16, 32, 64$ and $15\leq L_z\leq 256$ (all lengths in units
of $l_{max}$) for systems from $\Ntot=8192$ up to $32768$ monomers.

\section{KINETICS OF BRUSH FORMATION}
\label{sec:Kinetics}

Since the present work has been focused mainly on the equilibrium
properties of in situ grown polymer brushes (Sec.\ref{sec:Static}), 
we briefly report here some salient features of the kinetics of growth 
by reversible polymerization at constant \Ntot.
Some idea about the the conformation of the growing pseudo-brush may be 
gained from the series of snapshots of the polymer layer, Fig.\ref{fig:snap}, 
taken at successive times where the average chain length\cite{foot3} 
$6\le\Nav\le 20$.

At the start of the polymerization dimers ($N=2$-chains) are predominantly 
created with a large number of unsaturated initiators present.  
Each newly attached monomer becomes then an active chain end itself
(dark spheres in Fig.\ref{fig:sketchPB}), and may attach other free monomers 
in its vicinity or detach itself from the chain.
The snapshot on Fig.\ref{fig:snap} corresponds to a configuration 
at equilibrium. 
Evidently, while the density of the brush immediately at the grafting
surface is very high, the space above it is dominated by several very
long chains which barely overlap, so-called {\em mushrooms}. 
While shorter chains appear to be somewhat stretched this is definitely 
not true for the long chains (see Sec.\ref{subsec:Conformation}).
 
\begin{figure}
\centerline{\epsfig{file=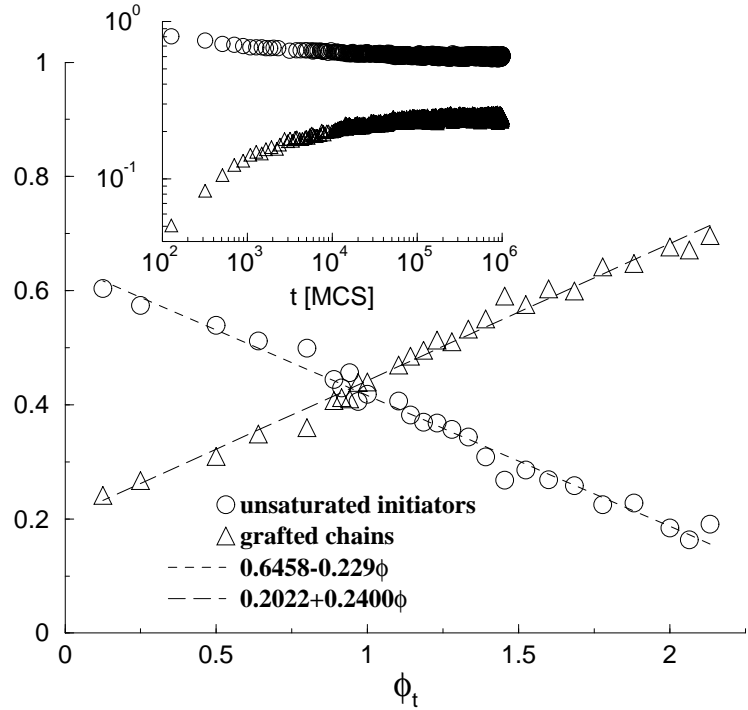,width=120mm,height=100mm}}
\caption{\label{fig:starter}
Activation of initiators at fixed total particle number $\Ntot=8192$: 
Fraction of unsaturated initiators ($N=1$: spheres) 
and of grafted chains ($N>2$: triangles) 
versus $\phitot \propto 1/\Lz$ (main figure).
In the inset the same quantities are traced
versus time elapsed after starting the polymerization reaction
for the standard system with $\Lz=256$ and $\phitot=0.125$.
}\
\end{figure}

Since the chemical reaction in our closed system proceeds at the expense
of the available free monomers, it is clear that the observed kinetics of 
polymerization should differ from that studied at constant incident 
flux\cite{WCJT96}. Thus no clear cut scaling relationships have been 
observed with the time evolution of the quantities of interest, 
as for example, in the case of surface coverage with time shown
in the inset of  Fig.\ref{fig:starter}. We show there the fraction
of grafted chains which are larger than $N=2$ and the fraction
of unsaturated initiators (i.e. chains of length $N=1$) versus time.
(The data has been gathered during runs of more than $10^6$ MCS and 
was averaged over $20$ such runs each starting with different 
initial configuration.)

We stress that the equilibration kinetics is rather slow compared to 
similar systems of GM in the bulk\cite{WMC98} 
which are allowed to break and where the dynamics is not limited to 
end monomer adsorption and desorption. This forced us to simulate
relatively short chains compared to our previous study \cite{MWL99b}.

The total number of monomers in the aggregate is initially diffusion-limited,
i.e. $\Ng \propto \sqrt{t}$, and then levels off in a non-trivial way 
(not shown). Similar graphs revealing no clear cut dynamic scaling have 
also been obtained for the time evolution of the mean chain length \Nav\
and the characteristic brush height \Hav\ .  
In contrast, $\la H(t) \ra$ plotted against 
\Nav\ during the relaxation of the system to equilibrium shows a nearly 
perfect relationship as $\Hav \propto \Nav^{1.425\pm 0.023}$ (not shown). 

\begin{figure}
\centerline{\epsfig{file=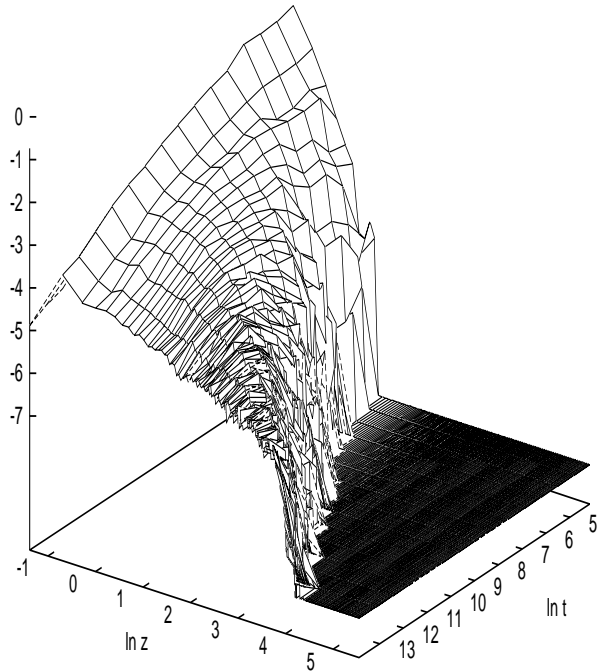,width=120mm,height=100mm,angle=270}}
\caption{\label{fig:phi_zt}
The density of grafted monomers $\phi(z,t)$ in a system of total concentration
$\phitot=0.125$ with $\Lz=256$ during relaxation to equilibrium. 
The resulting effective slope at late times is $\alpha \approx 2/3$.}\
\end{figure}

Eventually, in order to show that we have also reached equilibrium for
the distributions, we show in Fig.\ref{fig:phi_zt} 
the gradual formation of the typical density profile $\phi(z)$ with time $t$ 
elapsed after the beginning of the polymerization. 
We see that gradually the power law profile $\phi(z)\propto z^{-\alpha}$ 
(plus exponential cut-off) is build up. 
As will be discussed below (see Sec.\ref{subsec:Densities}) 
the exponent is consistent with $\alpha \approx 2/3$.

\section{EQUILIBRIUM PROPERTIES OF LP-BRUSHES}
\label{sec:Static}

Most of the equilibrium properties of the polymer brush are studied here
with varying total monomer concentration \phitot\ of the solution 
whereby the scission energy $J=2$ (and other bonded and non-bonded
energy parameters), the total number of monomers \Ntot\ in the box 
are kept constant as well as $\Lx=\Ly$,
whereas the box size in $z$-dimension \Lz \ is varied. 
We begin by analyzing the role of the box size on 
the fraction of grafted chains, the mean chain length \Nav\
and the force \Fav\ measured on the opposite wall,
before we proceed to analyze the MWD, the conformational properties
and the density profiles in view of the theoretical questions
formulated in Sec.\ref{sec:Theory}.

\subsection{Effects of box size variation}
\label{sec:Box}

We first examine the variation of the fraction of grafted LP chains
(with $N>2$) and the fraction of remaining unsaturated initiators
with system density. As depicted in Fig.\ref{fig:starter} both 
are found to change {\em linearly} with \phitot.
Note that these fractions do not add up to unity since there is always an 
important fraction of sites with only one monomer attached 
--- such $N=2$-chains are not counted as proper tethered polymers.

\begin{figure}
\centerline{\epsfig{file=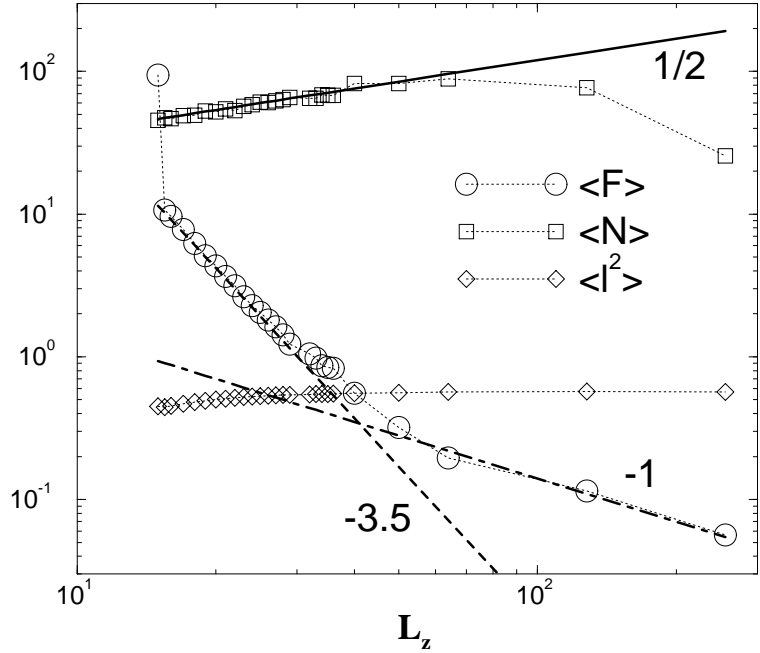,width=120mm,height=100mm}}
\caption{\label{fig:FNl_Lz}
Effects of the box volume $\propto \Lz$ at $\Ntot=8192$ and $L_x=L_y=16$.
The mean chain length \Nav \ (squares),
the brush pressure on the opposite wall \Fav\ (spheres)
and the average squared bond length between neighboring monomers
along the backbone of the polymer chains (diamonds)
are plotted as function of \Lz.
For large boxes $\Lz > 80$ the pressure is only due to the osmotic pressure 
of the free monomers and \Fav\ decreases with the volume, 
i.e. with slope $-1$. 
The bond length becomes constant.
For smaller boxes where the brush is directly interacting with the opposite 
wall we find $\Nav \propto \Lz^{0.5}$ (full line) and 
$\Fav \propto \Lz^{-3.5}$ (dashed line). 
}\
\end{figure}

The mean polymer chain length \Nav\
(i.e. the average degree of polymerization), 
the mean pressure \Fav\ on the opposite wall at $z=\Lz$ and 
the mean squared bond length $\la l^2 \ra$ as a function of \Lz  
are depicted in Fig.\ref{fig:FNl_Lz}. 
For small boxes where $\Lz < 80$ the LP brush interacts directly
with the opposite wall while for larger boxes this interaction
is only mediated via the ambient free monomer pressure.

In the small box limit \Nav\ {\em grows} with \Lz\ 
following a power-law relationship $\Nav \propto \phitot^{-1/2}$  
(full line in Fig.\ref{fig:FNl_Lz}). This surprising result clearly underlines the
difference of the present system with respect to living polymers in the bulk
where $\Nav \propto \phitot^{1/2}$ \cite{CC90,WMC98}. 
Evidently this effect is due to the influence of the upper
wall at distance \Lz\ on the polydisperse brush: 
as soon as \Lz\ becomes sufficiently large the monomers begin to dissociate 
from the active chain ends and the degree of polymerization drops.

The compression of the LP brush due to increasing confinement 
can also be clearly seen from the mean square of the bond length.
It is evident that a steady shrinking of the elastic bonds with density
takes place for $\Lz < 80$.

In order to characterize the dependence of the pressure within the box
on the total monomer concentration \phitot, 
i.e. to obtain an equation of state,
we have measured the force \Fav, exerted by the free monomers and the LP layer, 
on the top of the box. In order to measure the force exerted by the monomers 
on the upper wall directly, the wall has been supplied with
a Morse potential, eq.(\ref{eq:Morse}), at $z=\Lz$ where we have kept the 
repulsive branch only.
As expected, one observes a steadily decreasing \Fav\ vs \Lz\ relationship 
with growing distance \Lz. For $\Lz > 80$ the total pressure
on the opposite wall is mainly due to osmotic pressure of 
the free monomers in the solution, hence is proportional to the
free monomer density $u(z \gg H) \propto 1/\Lz$. 
(See Fig.\ref{fig:rhou}.) 
In contrast, for $\Lz < 16$ the system attains its maximum density of 
$\phitot \approx 2.13$ and 
the pressure rapidly grows as the hard core repulsion 
between the monomers comes into play. 
In between these extremes, however, there is an interval of heights \Lz\ 
where the pressure is found to follow a
power-law dependence on \Lz\ with an exponent of $\approx -3.5$ 
(long dashed line).
For these values of \Lz\ the opposing wall is in immediate contact with the 
polymer layer.

\subsection{Molecular Weight Distribution}
\label{subsec:MWD}

\begin{figure}
\centerline{\epsfig{file=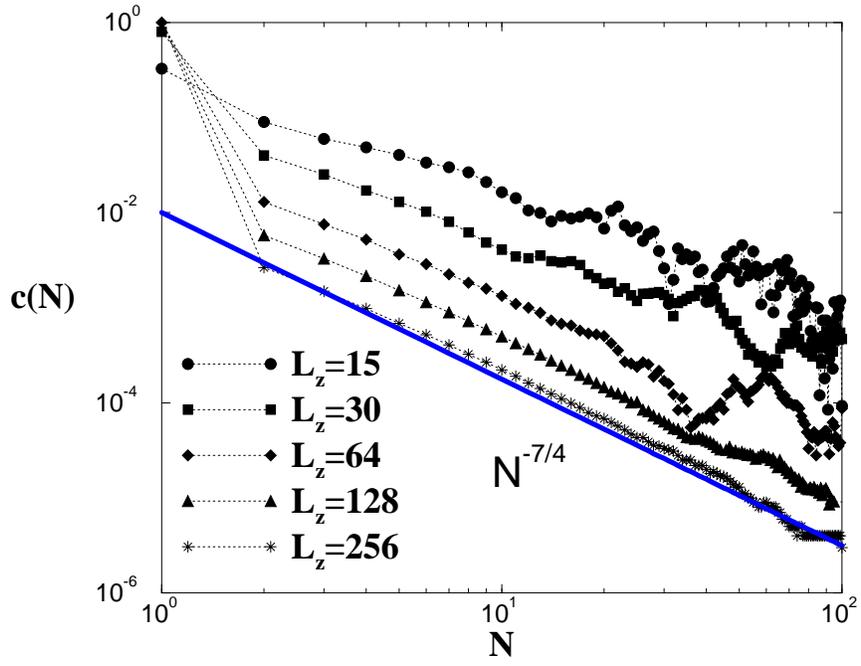,width=120mm,height=100mm}}
\caption{\label{fig:mwd}
MWD of a dense polydisperse brush vs $N$ for different total 
concentrations $\propto 1/\Lz$ with $\Ntot=8192$  
showing clearly a power law $c(N) \propto N^{-\tau}$ with $\tau \approx 7/4$.
This is the central result of this paper.
}\
\end{figure}

In Fig.\ref{fig:mwd} we present the observed MWD $c(N)$  
(normalized to the mean number of chains of length $N$)
of grafted LP chains versus chain length $N$.
Systems with $16\times 16 \times \Lz$ containing $8192$ particles are 
considered here. 
Despite some scatter of data for the longer chains at higher concentrations, 
a power-law relationship $c(N)\propto N^{-\tau}$ with $\tau \approx 7/4$ 
is evident. 
Hence, the MWD of grafted LP decays indeed much slower than the
exponential behavior found in the corresponding LP bulk systems
briefly mentioned in Sec.(\ref{subsec:MFassumption}). Evidently,  
this is due to the monomer density gradient within the dense pseudo-brush
favoring the longer chains and is to be expected.
Much more surprising is the value of the exponent found.
It is clearly smaller than the exponent $\tau=11/5$ expected for a 
mushroom-like layer and is (within numerical accuracy) identical to 
the prediction of the DLAWB problem described in Sec.\ref{subsec:DLAWB}.
This finding clearly disagrees with the SCF approach presented in 
Sec.\ref{subsec:SCF_LPBRUSH} where we estimated the free energy of the layer
by counting the numbers of excluded volume blobs. The layer appears to be
much more brush-like. This is the central result of the paper.

We do not have yet any completely satisfactory explanation for the observed
$\tau$. Presumably this is caused by logarithmic corrections which 
lower the free energy with respect to our steepest-decent estimation
of Sec.\ref{subsec:SCF_LPBRUSH}.
The fact that the MWD is of power-law form shows that they scale like
the number of blob term and cannot be neglected.

\begin{figure}
\centerline{\epsfig{file=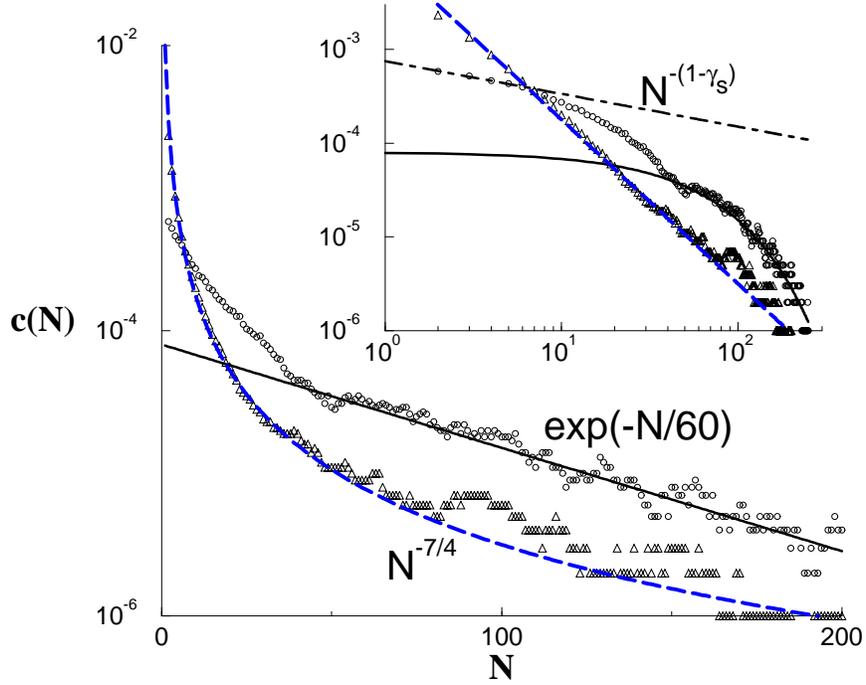,width=120mm,height=100mm}}
\caption{\label{fig:mwd_w}
MWD $c(N)$ for two very large systems 
(both of total density $\phitot=0.125$ containing $\Ntot=32768$ particles) 
with high ($d=1$ : triangles) and low ($d=4$ : spheres) grafting density
of the initiators.
The first system confirms the power law behavior with $\tau \approx 7/4$
(dashed line) while the mushroom-like system shows for large $N$ essentially 
an exponential decay (full line). However, as can be seen from the
inset where we have retraced the data in log-log coordinates,
this MWD is not incompatible with the weak initial power law behavior
(dashed-dotted line) expected for non-overlapping mushrooms
fixed on an impenetrable wall.
}\
\end{figure}

Finite size effects in the present simulation are negligible. 
This is clearly shown in Fig.\ref{fig:mwd_w} where a much larger system of 
$32768$ particles in a box of $32\times 32 \times 256$
with densely ($d=1$) placed initiators on the surface 
reveals the same power exponent $\tau=7/4$, as in Fig.\ref{fig:mwd}. 

If the distance between anchoring sites $d$ is further
increased and the chains become too small (compared to $d$) to overlap,
we expect the distribution eq.(\ref{eq:cNmush}) for mushrooms grafted
on an impenetrable surface with a weak power law 
$c(N) \propto N^{-(1-\gamma_s)}$ with $\gamma_s\approx 0.65$ 
and a pronounced exponential tail for larger $N$.
This is clearly shown in Fig.\ref{fig:mwd_w} (spheres) for the 
system $64\times 64 \times 64$ (i.e. of same total density $\phitot=0.125$)
with large distance $d=4$ between chain anchoring initiators. 
In the inset we test the power law decay anticipated for small $N$ 
due to the effective repulsion of the hard wall.
Qualitatively, this is confirmed (dashed-dotted line).
Because the chains within this regime are not very long one could not expect
to obtain a better estimate. Essentially much larger \Nav\ und $d$ are needed
than we are at present able to simulate. 
Note that similar (finite-size related) problems have also been reported for 
the estimation of the $\gamma$ exponent from the MWD of reversible 
polymerization in the bulk\cite{WMC98}.

\subsection{Conformational Properties}
\label{subsec:Conformation}

\begin{figure}
\centerline{\epsfig{file=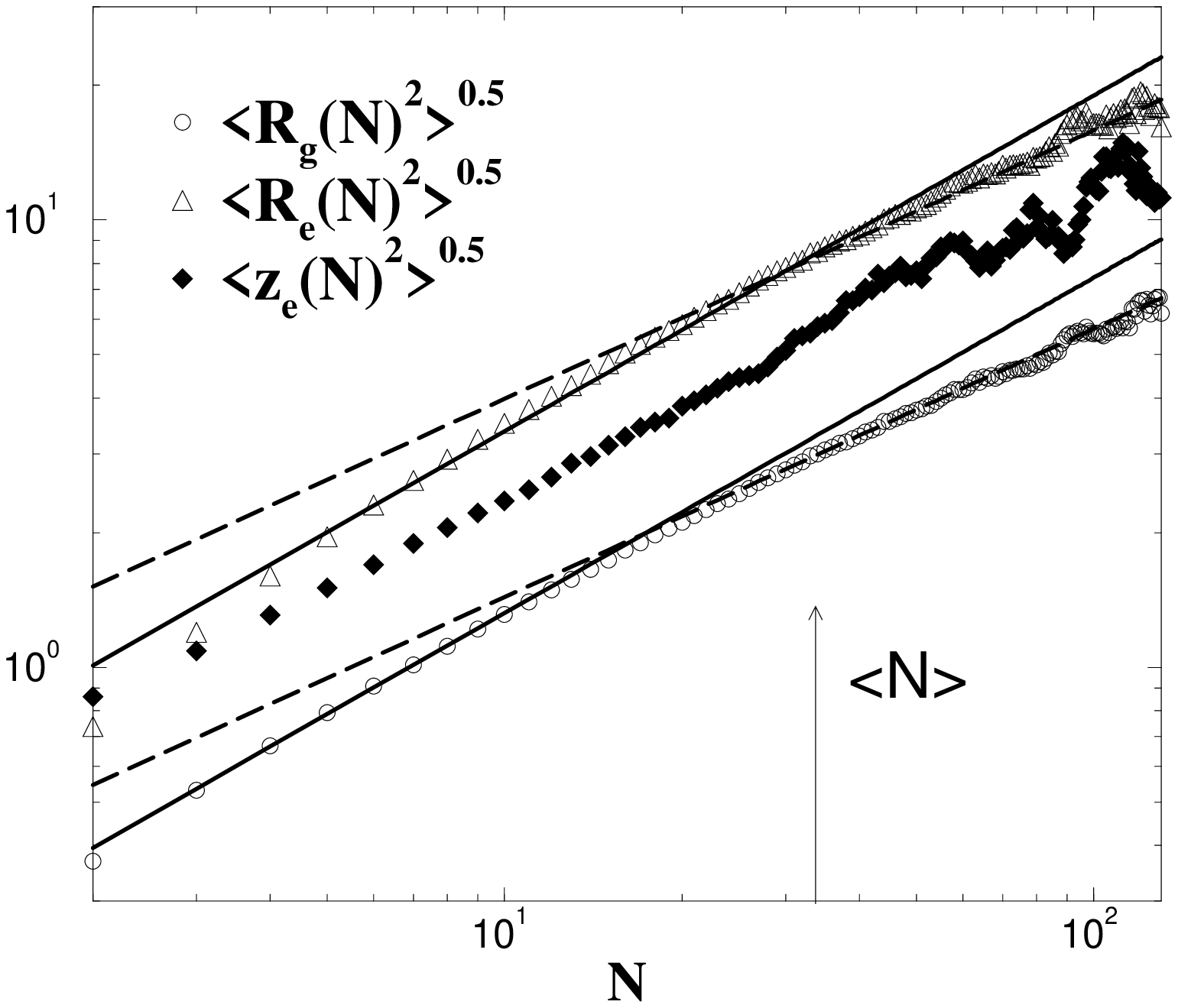,width=120mm,height=100mm}}
\caption{\label{fig:R_N}
Radius of gyration $\Rgg^{0.5}$ (spheres), 
end-to-end distance $\Ree^{0.5}$ (triangles)
and mean squared end monomer position $\zee^{0.5}$ (full diamonds)
averaged over chains of given mass $N$. 
Note height $\Hav=18.2$ and mean chain length $\Nav=33.9$ of this 
configuration at $\phitot=0.125$.
Dashed lines denote the slope of $\nuperp = 0.6$
corresponding to unstretched swollen coils in the good-solvent limit. 
The full lines indicate the slope $\nuperp = 0.75$ consistent with a
strongly stretched LP brush characterized by the exponent $\tau=7/4$.}\
\end{figure}

The observed power-law behavior of the MWD with $\tau=7/4$
does not necessarily mean that we have a dense polydisperse
brush in the strong-stretching limit. 
In order to check this assumption directly we investigate
the mean squared end-to-end distance \Ree, 
the radius of gyration \Rgg\
and the mean squared vertical position of the end monomer \zee.
These quantities are plotted in Fig.\ref{fig:R_N}
as functions of the chain length $N$.
The straight lines with slope $\nuperp = 0.6$ demonstrate that 
the few very long and, hence, non-overlapping chains ($N > \Nav$) behave, 
as one expects, like flexible coils in a good solvent. 
More interestingly, though, the shorter chains show a steeper 
increase with exponent $\nuperp \approx 0.75$.
(This is clearly seen from the end-to-end distance and the radius of 
gyration and to a lesser degree from \zee.)
The exponent $\nuperp=3/4$ is expected from eq.(\ref{eq:nuperp_tau}) for a 
strongly stretched layer for a MWD described by exponent $\tau=7/4$.
Hence, the slopes of Fig.\ref{fig:R_N} and Fig.\ref{fig:mwd},
and the strong-stretching assumption, are consistent.
Consequently, the new exponent coincides again with the DLAWB prediction,
but not with our SCF approach.

\begin{figure}
\centerline{\epsfig{file=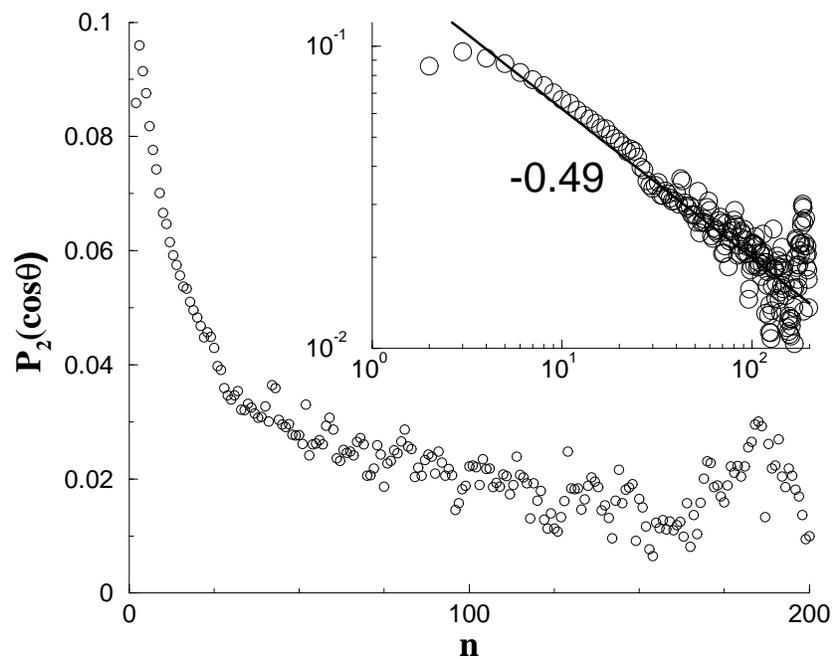,width=120mm,height=100mm}}
\caption{\label{fig:kos_n}
Variation of the "order parameter" of bond orientation, $P_2(\cos(\theta))$,
with bond index $n$. The data were obtained for a large box
$32 \times32 \times 256$ containing $32768$ monomers.
The same data is depicted in the inset in logarithmic coordinates 
and compared with the power law fit $P_2(n)\propto n^{-0.49}$.}\
\end{figure}

The stretching of the polymer chains in the LP brush can be also made 
visible by looking at the second Legendre polynomial
$P_2(\cos(\theta))=[3\langle \cos(\theta)\rangle -1]/2$ 
as a measure for the average orientation of the covalent bonds with respect 
to the $z$-axis, normal to the grafting surface. 
Here $\theta$ measures the angle between the bond
vector and the $z$-axis, normal to the grafting surface.
One would find $P_2 = - 0.5$ for bonds lying flat on the surface, 
$P_2 = 1$ for bonds perpendicular to it, and $P_2 = 0$ if all
the bonds are oriented completely at random in the $3D$-space. 
In Fig.\ref{fig:kos_n} we plot $P_2$ versus the bond index $n$
(between the monomer $n$ and $n+1$ where the monomers are counted
along the chain starting with the initiator $n=1$) 
averaged over all chains (with $n \leq N$). 
One can see that the bonds between neighbor monomers stay predominantly
vertical to the surface, and the smaller $n$, i.e. the closer the bond
is to the surface, the more pronounced this orientation is.
With increasing bond index $n$ and, hence, decreasing monomer density
$\phi(z(n))$, the local orientations become more and more unrestricted by 
the presence of other chains. As shown in the inset, the data compares well 
with the power law fit $P_2(n) \propto n^{-0.49}$. This underlines again
the power law character of the LP brush.

\subsection{Density Profiles}
\label{subsec:Densities}

From the observed MWD and the strong-stretching behavior discussed above
we may now anticipate power law decay for both the equilibrium density profile 
of end monomers $\rho(z) \propto z^{-\beta}$ and 
of grafted monomers $\phi(z) \propto z^{-\alpha}$.
Specifically, from $\tau \approx 7/4$ one expects
$\beta \approx 2$ and $\alpha \approx 2/3$ 
where we have used eq.(\ref{eq:beta_tau}) and eq.(\ref{eq:alpha_tau}) 
respectively. (Obviously, the same predictions could have been made
starting from $\nuperp=3/4$.)

\begin{figure}
\centerline{\epsfig{file=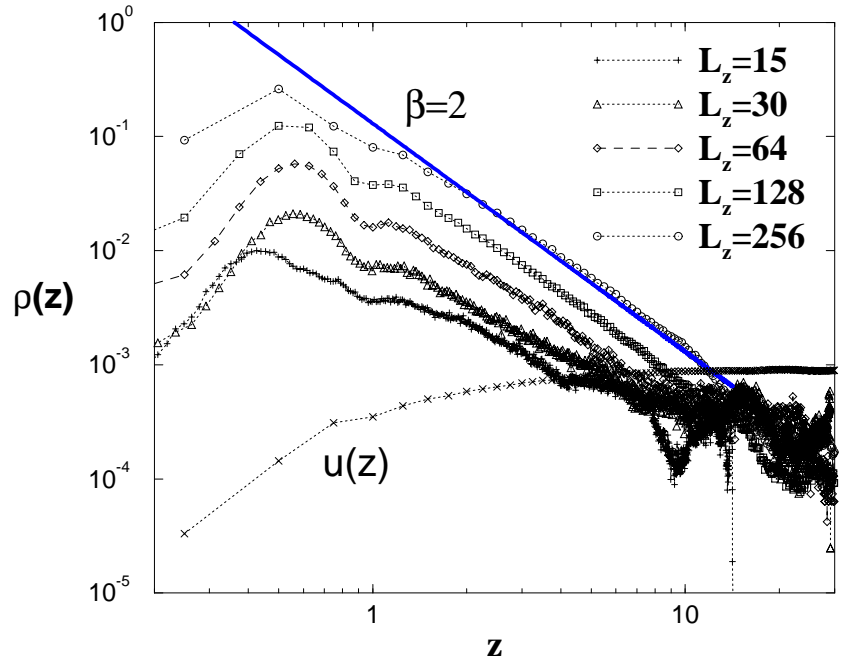,width=120mm,height=100mm}}
\caption{\label{fig:rhou}
Density profiles for the active end-monomers, $\rho(z)$, 
in systems of $\Ntot=8192$ particles within a $16\times 16\times \Lz$ box.
The density of free monomers $u(z)$ (stars) at $\Lz = 256$ is included as well. 
The full line denotes the exponent $\beta=2$ expected from the MWD and
the strong-stretching assumption.
}\
\end{figure}

\begin{figure}
\centerline{\epsfig{file=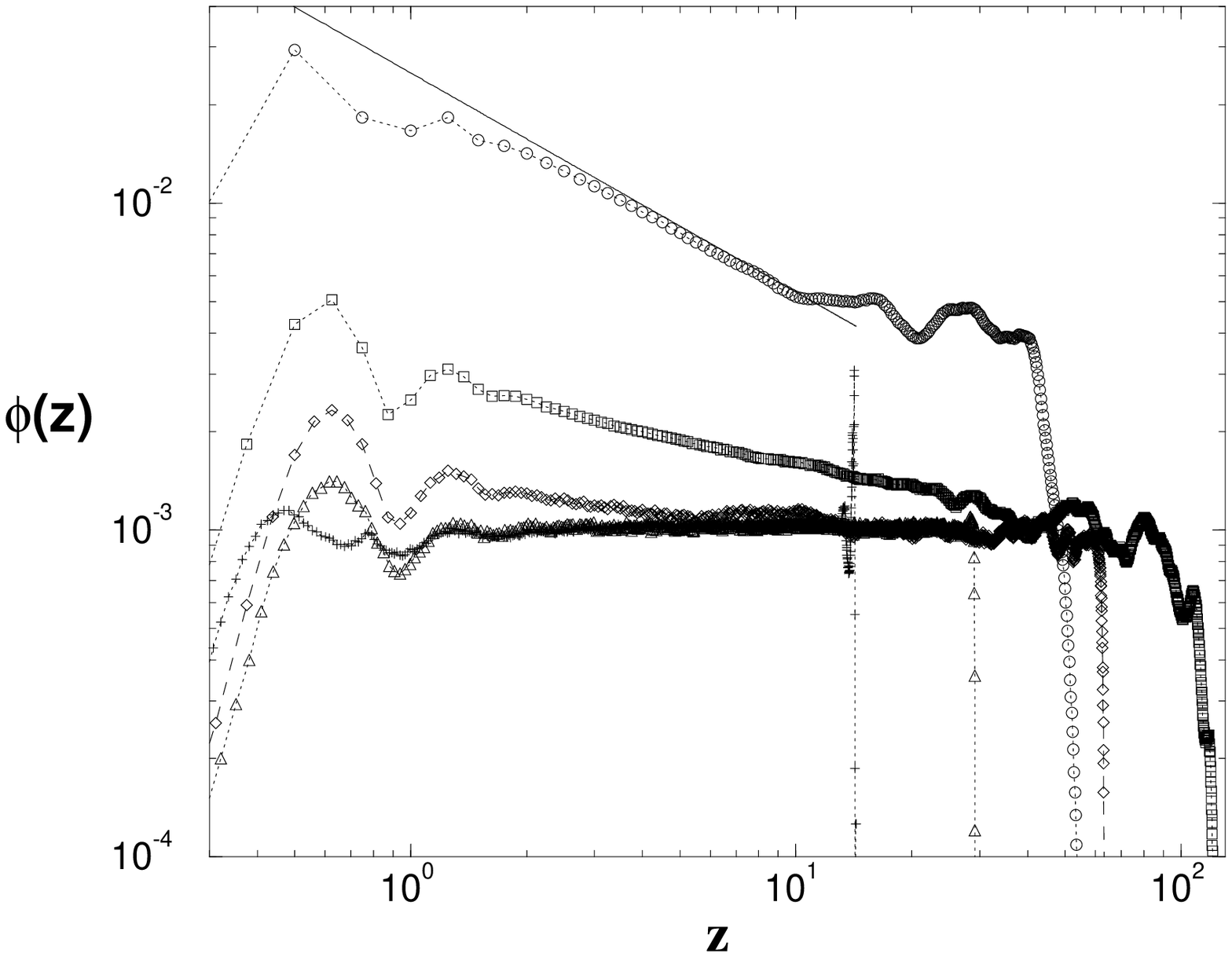,width=120mm,height=100mm,angle=0}}
\caption{\label{fig:phi}
Density profiles $\phi(z)$ of aggregated monomers vs distance from 
the grafting surface. 
Same symbols and configurations as in Fig.\protect\ref{fig:rhou}.
The full line denotes a slope of $\alpha=2/3$
consistent with eq.(\ref{eq:alpha_tau}) and $\tau=7/4$.}
\end{figure}

This compares surprisingly well with the slopes shown in Fig.\ref{fig:rhou} 
and Fig.\ref{fig:phi} for the dilute limit with $\Lz=256$. 
(Note that the $\alpha$-value is slightly too large.)
Hence, the observed set of four exponents 
$\tau$, $\nuperp$, $\beta$ and $\alpha$ is self-consistent.

Here we have presented results for different concentrations with $\Ntot=8192$ 
particles in a box of size $16\times 16\times \Lz$. 
All profiles are normalized to unity, e.g.  $\int_0^{L_z}\phi(z)dz = 1$.
We observe the same values of the exponents $\alpha$ and $\beta$ 
when the total concentration is further decreased, i.e. \Lz\ increased. 
Even at higher densities, up to $\phitot = 2.0$ (at $\Lz = 16$), 
both $\phi(z)$ and $\rho(z)$ retain their power-law dependence on $z$. 
While the function $\rho(z)$ remains largely unaffected, 
the effective exponent $\alpha$ decreases.
This is expected qualitatively for a brush compressed between two walls
where at very high compression it becomes a melt of 
blobs of constant size $\xi$.

In Fig.\ref{fig:rhou} we have also included the density profile of the 
free monomers in equilibrium, $u(z)$. It becomes evidently constant 
for large $z$ where the volume fraction of grafted monomers becomes negligible.
Obviously, the chemical potential of the free monomers has to be constant with 
respect to position, i.e.  $\mu = \mu_{ex}(z) + \log(u(z)) = const$. 
Within mean-field one expects an excess chemical potential of
$\mu_{ex}(z) \propto \phi(z)+u(z)$ which is $\approx \phi(z)$ for $z \ll H$. 
This is indeed born out as one verifies by plotting $\log(u(z))$ versus $\phi(z)$
(not shown). This explains the slightly more complicated (not power law) form
of the observed $u(z)$.

\section{DISCUSSION}
\label{sec:Discussion}

To our knowledge, this is the first time that this problem (as defined in
Fig.\ref{fig:sketchPB}) was considered by numerical as well as analytical 
means and, admittedly, our understanding of the LP brush is still incomplete.

The {\em effective} exponents $\tau=7/4$, $\nuperp=3/4$, $\beta=2$ and 
$\alpha=2/3$ found in the simulations form a {\em self-consistent} set of exponents 
in agreement with the strong-stretching predictions 
eqs.(\ref{eq:nuperp_tau},\ref{eq:beta_tau},\ref{eq:alpha_tau}). 
Surprisingly, these values reveal a nearly perfect agreement (within
inevitable statistical error) with the 
predicted exponents\cite{WCJT96} for the apparently different problem 
of the {\em in situ} formation of a grafted polymer layer 
through diffusive-limited aggregation with {\em irreversible} polymerization
(DLAWB).
It is important to stress, however, that our values might still be somewhat 
off the DLAWB exponents due to the additional exponential cut-off 
(effecting the long chains) and the too short mean chain lengths \Nav\
(hence, the too small number of decades) we have been able to simulate.

Yet the question about the observed similarity between a pseudo-brush with an
{\em annealed} MWD in thermal equilibrium and a steady-state DLAWB
brush with a {\em quenched} MWD remains.
In order to understand these results better we have attempted to describe
the layer by a standard self-consistent field theory (SCF). 
As a steepest-decent estimation of the free energy of our polydisperse LP 
brush we used the number of blobs contained in the layer ignoring all
logarithmic corrections \cite{Godefroid}. This yields a (marginally 
self-consistent) self-similar mushroom-like layer with $\xi(z) \propto z$.
Hence, every chain should be essentially contained in its own blob. 
However, the resulting fluffy layer is inconsistent with the 
strongly overlapping pseudo-brushes with $\xi(z) \propto \sqrt{z}$ 
we find in the computer experiment.
In particular, the observed decay of the MWD with $\tau\approx 7/4$
is much slower than the SCF result $\tau = 11/5$.
This is the surprising and central result of this paper.
It clearly shows that the free energy of the brush in our approach has been
overestimated and that the logarithmic corrections have to be taken  
properly into account \cite{Godefroid}. Unfortunately, these corrections are 
not well mastered even for monodisperse quenched polymers \cite{NS98}.
At best one may take at present $\tau$ as a phenomenological parameter 
characterizing the free energy of the layer.
It would be interesting to estimate numerically the free energy of the layer
by some direct method by varying the interaction parameters, although
we have not attempted this yet.
Thus further studies, numerical as well as analytical, would be required 
for a full insight into the equilibrium behavior of the LP brush.

\section{SUMMARY}
\label{sec:Summary}
Polydisperse brushes of high grafting density, obtained by {\em reversible}
radical chain polymerization reaction onto a solid substrate with
surface-attached initiators, have been studied by means of an off-lattice 
Monte Carlo model of living polymers (LP). 
Basically, our simulational results for the layer in equilibrium
indicate that a moderately stretched pseudo-brush is formed which is
well described by diverse power law distributions
plus an additional exponential cut-off at the edge of the layer:
\begin{itemize}
\item
The Molecular Weight Distribution (MWD) of the chains of contour length $N$
fits $c(N)\propto N^{-\tau}$ with $\tau \approx 7/4$. Hence, the MWD of
grafted LP decays much slower than corresponding LP bulk systems.
This is due to the monomer density gradient within the dense pseudo-brush 
favoring the longer chains.
\item
If the chain overlap becomes small the MWD becomes essentially exponential
with a weak singularity for small $N$ with $\tau = 1-\gamma_s \approx 0.35$
caused by the effective repulsion of the impenetrable wall.
\item
The end-to-end distance and the radius of gyration show
$R \propto N^{\nuperp}$ with $\nuperp \approx 3/4$ for smaller chains
and $\nuperp \approx 0.6$ for long enough, not overlapping mushrooms.
\item
The density of active chain ends is well fitted by 
$\rho (z)$ $\propto z^{-\beta}$ with $\beta \approx 2$ within a broad
concentration regime.
The density profile of aggregated monomers scales roughly like
$\phi (z)\propto z^{-\alpha}$ where the exponent becomes 
$\alpha \approx 2/3$ in the dilute limit.
\end{itemize}

Additionally, the influence of total monomer concentration
and of grafting density of the initiators on the density profiles, on the
average degree of polymerization, and on the conformational properties of
the polymer chains in the brush have been studied too.
We have also investigated the force, exerted by the brush on the opposing
wall of the container. Very briefly, we have also examined the kinetics of
brush growth until the onset of equilibrium.
 
\section{ACKNOWLEDGMENTS}
This research has been supported by the National Science Foundation,
Grant No. INT-9304562 and No. DMR-727714, and by the Bulgarian National
Foundation for Science and Research under Grant No. X-644/1996.
JPW thanks S. Godefroid, A.Tanguy and I.Paganabarraga 
for stimulating discussions. The authors thank J. L. Barrat for critical 
reading of the manuscript.

\newpage
\begin{figure}
\caption{\label{fig:snap}
A snapshot of a brush configuration
corresponding to mean chain length $\Nav \approx 20$.
Neither the ambient free monomers nor the array of initiators are shown. 
Different intensity of grey color applies to different chains.
Only the shorter chains do sufficiently overlap and, hence,
are somewhat stretched by the density gradient.}\
\end{figure}

\newpage

\newpage

\newpage

\newpage

\newpage

\newpage

\newpage

\newpage

\newpage


\begin{references}

\bibitem{HTL92}  A. Halperin, M. Tirrell, T. P. Lodge, 
Adv. Polym. Sci. {\bf 100}, 31(1992).

\bibitem{BLOBBRUSH}  
S. Alexander, J. Physique, {\bf 38}, 983(1977);
P.G. De Gennes, Macromolecules, {\bf 13}, 1069 (1980).

\bibitem{SCFBRUSH} A.Semenov, JETP, {\bf 88}, 1242(1986);
S. T. Milner, T. A. Witten and M. E. Cates,
Macromolecules, {\bf 21}, 2610 (1988).

\bibitem{NS98}
R. R. Netz and M. Schick, Macromolecules, {\bf 31}, 5105 (1998).

\bibitem{Slowprocess} 
A. Johner and J. F. Joanny, J. Chem. Phys. {\bf 98}, 1647 (1993);
J. Wittmer, A. Johner, J.F. Joanny, K. Binder,
J. Chem. Phys., {\bf 101} 4379 (1994). 

\bibitem{Ruehe}  O. Prucker and J. R\"{u}he, Macromolecules, {\bf 31},
602(1998).

\bibitem{WCJT96}
J. P. Wittmer, M. E. Cates, A. Johner, and M. S. Turner,
Europhys. Lett., {\bf 33}, 397 (1996)


\bibitem{DLA} T. A. Witten, L. M. Sander, Phys. Rev. B {\bf 27}, 5686(1983).

\bibitem{Guis}  O. Guiselin, Europhys. Lett. {\bf 17}, 225(1992).

\bibitem{Needles}  
P. Meakin, Phys. Rev. A {\bf 33}, 3371(1986);
G. Rossi, Phys. Rev. {\bf A35}, 2246(1987);
M. E. Cates, Phys. Rev. A {\bf 36}, 5007(1986);
J. Krug, K. Kassner, P. Meakin, and F. Family, 
Europhys. Lett. {\bf 24}, 527(1993).

\bibitem{Degennesbook}
P.G. De Gennes, {\em Scaling Concepts in Polymer Physics}
(Cornell University Press, Ithaca, NY, 1979).

\bibitem{DJbook} 
J. Des Cloizeaux and G. Jannick, {\em Polymers in Solution},
Clarendon Oxford (1990).

\bibitem{Greer} S.C. Greer, Advances in Chemical Physics {\em 96}, 261 (1996).

\bibitem{CC90}
M. E. Cates and S. J. Candau, J. Phys. Cond. Matt. {\bf 2}, 6869(1990).

\bibitem{WMC98}
J. P. Wittmer, A. Milchev, and M. E. Cates,
J.Chem. Phys. {\bf 109}, 834(1998); Europhys. Lett. {\bf 41}, 291(1998).

\bibitem{Godefroid}
S. Godefroid, {\em Etude statistique de la polymerization in situ d'une
brosse}, Master thesis, Ecole Normale Superieure de Lyon, Lyon, France (1999).

\bibitem{MWL99b}
A. Milchev, J.P. Wittmer, D. Landau, 
in preparation.

\bibitem{MPB} A. Milchev, W. Paul and K. Binder, 
J. Chem. Phys. {\bf 99}, 4786(1993).

\bibitem{AMKB} A. Milchev and K. Binder, 
Macromolecules, {\bf 29}, 343(1996). 

\bibitem{foot1} In principle an additional Lagrange multiplier might
be introduced to fix the density of grafted chains.
One can, however, easily show that this parameter can always
be incorporated in the scission energy $J$ which penalizes
every chain end.

\bibitem{foot2} We note for completeness and later reference that the 
mean-field eq.(\ref{eq:cN}) is very successful for dense LP in the bulk
as well\cite{WMC98,MWL99b}. ($c(N)$ here denotes then a volume density.)
There $F_{chain}$ is a constant with respect to $N$, but a function of the 
local density, and the MWD is again a pure exponential.

\bibitem{foot3} We define here the mean chain length as
$\Nav = \sum_{N>1} N c(N)/\sum_{N>1} c(N)$, $c(N)$ being the MWD,
i.e. the sum does not include the large number of free monomers.
Note that $N$ includes the initiator, i.e. $N=1$ for a chain
consisting of one initiator only.
\end{references}
\end{document}